# Towards Designing Systems with Large Number of Antennas for Range Extension in Ground-to-Air Communications

Haneya Naeem Qureshi, Ali Imran
School of Electrical and Computer Engineering, University of Oklahoma, Tulsa, USA
Emails: haneya@ou.edu, ali.imran@ou.edu

*Abstract*—Providing broadband connectivity to airborne systems using ground based cellular networks is a promising solution as it offers several advantages over satellite-based solutions. However, limited range of terrestrial base stations is a key challenge in full realization of this approach. This paper addresses this problem by proposing a mathematical framework for range extension leveraging large number of antennas at the base station. In contrast to prior works where range is not considered as a design parameter, we model the signal to noise ratio as a function of both number of antennas as well as the range in line-of-sight ground-to-air systems. This allows us to derive analytical expressions to determine the number of antennas required to increase range in different frequency bands and tracking and non tracking scenarios.

## I. Introduction

Systems "above the clouds" [1] are the only remaining areas today where broadband services are not fully available. Driven by users' demand for seamless connectivity regardless of location and time, providing connectivity in airborne systems is indispensable to the future of aircraft industry. According to a recent survey, nearly 75% passengers are ready to switch airlines to have access to faster internet and more than 20% passengers have already switched their airline for the sake of better in-flight internet access [2]. Like current aircraft passengers, these high user expectations are equally anticipated for future users of 'flying cars' as well as in unmanned aerial vehicle applications in next generation systems. This challenge is likely to aggravate in the next few years with increase in air travel and smart devices carried by passengers.

To this end, solutions for air to ground communications using ground based cellular networks is the subject of several recent studies due to the multitude of advantages offered by this approach as compared to satellite-based connectivity [3]-[6]. While satellite communication has been used for voice communication, its intrinsic capacity limits, high latency, lack of scalability and cost make it unsuitable for carrying multimedia and realtime traffic [4]. On the other hand, utilizing a ground-based cellular system to create a direct link between the airborne systems and the ground for broadband connectivity is a fast, scalable and an economical solution. Additionally, unlike satellite based approach, terrestrial cellular approach allows expansion of the network capacity exactly where it is needed by adapting the cell sizes or increasing the number of cells [3].

However, limited range of ground-based cellular systems hinders the full realization of this approach. This problem is more pronounced in parts of the world, such as Europe, where land masses are separated by seas. In such cases, base stations can only be deployed at most at the edges of the land masses, thus limiting the range of service from ground to airborne system. In this paper, we propose a solution to this problem by leveraging large number of antennas at the base station. While systems with large number of antennas have been widely studied in terrestrial networks, the main focus has been on capacity enhancement such as in [7]-[8]. Range extension using large number of antennas is not considered in terrestrial systems due to the large number of multipaths. Therefore, Signal-to-Noise (SNR) is either computed at cell center, cell edge or by averaging over all user locations drawn from a fixed distribution. However, two recent studies [9]-[10] have focused on cell coverage extension in terrestrial systems. Authors in [9] propose a rate-based cell coverage expansion scheme in ground-based systems while authors in [10] leverage orthogonal random precoding in Massive MIMO terrestrial systems. On the other hand, we investigate the use of large number of antennas itself for range extension in ground-to-air communications where line-of-sight is the dominant path. The focus of this study is to investigate SNR as a function of both number of antennas as well as the range. To the best of authors' knowledge, such three dimensional analysis leading to an SNR expression as function of range, elevation angle and number of antennas does not exist in current literature. Our derived expressions also allow analysis of different scenarios that stem from change in operational fre-quency and tracking capability of the antenna system and thus provides an essential tool for dimensioning systems with large number of antennas for ground-to-air communications.

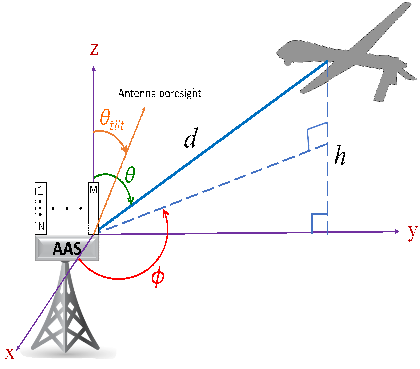

**Fig. 1:** System model.

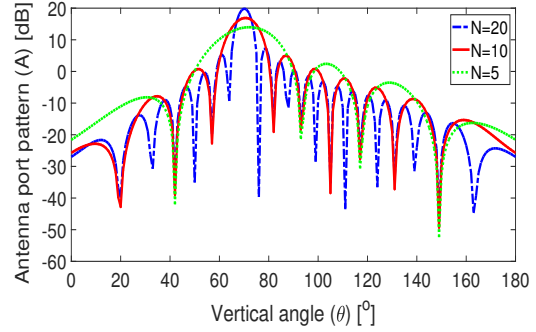

**Fig. 2:** Antenna port radiation pattern for $\theta_{tilt} = 70^o$.

This paper is organized as follows: The system model is described in Section II and a mathematical framework for range extension is presented in Section III. Section IV presents the numerical results and analysis and Section V concludes this paper.

## II. SYSTEM MODEL

We consider a UAV equipped with single receive antenna at a height of $h$ meters being served by an active antenna system (AAS) at the base station (BS) as shown in Fig 1. The range, which is distance from the center of the BS antenna array to the UAV is denoted by $d$.

### A. Antenna Array Model

We consider full dimension MIMO (FD-MIMO) utilizing an AAS with 2D planar antenna array structure. Unlike conventional LTE systems, for FD-MIMO, 3GPP proposes the organization of the radio resource implementation on the basis of antenna ports and antenna elements [11]-[12], where each column of active antenna elements in the array is referred to as an antenna port. We consider an $M \times N$ array comprising of $N$ elements and $M$ ports. Since all elements in the antenna array carry the same signal, each antenna port is perceived as a single antenna by the UAV.

The 3GPP proposed element radiation pattern is given by [12]:

$$A_E(\theta, \phi) = G_m - \min\{-(A_v(\theta) + A_h(\phi)), A_m\} \quad (1)$$

where $G_m$ is the maximum element gain, $\phi$ and $\theta$ denote the azimuth and vertical angles respectively. $\theta$ and angle of elevation of the BS to UAV are complementary angles. The radiation patterns in vertical and horizontal directions are modelled as:

$$A_v(\theta) = -\min\left[12\left(\frac{\theta - 90}{\theta_{3dB}}\right)^2, A_v\right] \quad (2)$$

$$A_h(\phi) = -\min\left[12\left(\frac{\phi}{\phi_{3dB}}\right)^2, A_m\right] \quad (3)$$

where $\phi_{3dB}$ and $\theta_{3dB}$ are the half power beamwidths in the azimuth and elevation domains respectively, $A_m$ is the maximum attenuation and $A_v$ is the vertical side lobe attenuation.

The array factor matrix for AAS, $\mathbf{A}_F$ is given by [12] as

$$\mathbf{A}_F = \mathbf{W} \circ \mathbf{V} \quad (4)$$

where $\circ$ is the Hadamard product and $\mathbf{V}$ and $\mathbf{W}$ are $N \times M$ matrices containing the array responses of individual radiation elements and weights to be applied to these elements respectively. Each entry of these matrices in the $r^{th}$ row and $c^{th}$ column is given as [11]:

$$v_{r,c} = \exp\left(i2\pi\left((c-1)\frac{d_h}{\lambda}\sin\phi\sin\theta + (r-1)\frac{d_v}{\lambda}\cos\theta\right)\right) \quad (5)$$

$$w_{r,c} = \frac{1}{\sqrt{NM}}\exp\left(-i2\pi\left((c-1)\frac{d_h}{\lambda}\sin\phi_{scan}\sin\theta_{tilt_c} + (r-1)\frac{d_v}{\lambda}\cos\theta_{tilt_c}\right)\right) \quad (6)$$

where $d_h$ is the horizontal separation between antenna ports, $d_v$ is the vertical separation between antenna elements, $\phi_{scan}$ is the horizontal steering angle and $\theta_{tilt_c}$ is the downtilt angle for the $c^{th}$ port.

The overall radiation pattern for $c^{th}$ antenna port can now be represented as:

$$A(\theta, \phi, \theta_{tilt}) = A_E(\theta, \phi) + 20\log_{10}|A_{F_c}(\theta, \theta_{tilt})| \quad (7)$$

where $A_{F_c}$ is the sum of entries in $c^{th}$ column of the matrix $\mathbf{A}_F$. In this work, we assume $\theta_{tilt_c} = \theta_{tilt}$. The antenna pattern is shown in Fig. 2 for $\theta_{tilt_c} = 70^o$, $\phi = 0^o$, $G_m = 8$dBi and $\phi_{3dB} = \theta_{3dB} = 65^o$. The main lobe shifts to the desired value of downtilt and becomes narrower with increasing antenna elements.

## B. 3D Channel

The channel gain between the transmitting antenna port and the receiver (UAV) can be represented as:

$$h_c = \sum_{(r \in \text{port } c)=1}^{N} w_{r,c}(\theta_{tilt}) \sqrt{G(\theta, \phi)} \, v_{r,c}(\theta, \phi) \quad (8)$$

$$= \mathbf{w}_c^T(\theta_{tilt}) \sqrt{G(\theta, \phi)} \mathbf{v}_c(\theta, \phi) \quad (9)$$

for $c = 1, \ldots, M$. $\mathbf{w_c}$ and $\mathbf{v_c}$ are the $c$-th columns of the matrices $\mathbf{W}$ and $\mathbf{V}$ respectively and $G$ represents the vertical and horizontal attenuation in linear scale given as:

$$G(\theta, \phi) = 10^{-1.2\left(\left(\frac{\phi}{\phi_{3dB}}\right)^2 + \left(\frac{\theta-90}{\theta_{3dB}}\right)^2\right)} \quad (10)$$

## C. Downlink SNR

The received complex baseband signal at the user is given by:

$$y = \sqrt{P_t F 10^{\frac{G_m}{10}}} \mathbf{h}^H \tilde{\mathbf{x}} + \mathbf{n} \quad (11)$$

where $P_t$ is the transmitted power, $G_m$ is the maximum antenna gain, $\mathbf{h} = [h_1 \ldots h_M]$ is the $1 \times M$ complex channel vector from the BS to the user given by (9), $\tilde{\mathbf{x}}$ is the precoded transmit signal from AAS, $n$ is the additive white Gaussian noise with zero mean and variance equal to $\sigma_n^2$ and $F$ is the free space path loss given by:

$$F = \left(\frac{\lambda}{4\pi d}\right)^2 \quad (12)$$

Assuming perfect channel state estimation at the transmitter and employing conjugate precoding at the transmitter, i.e., $\tilde{\mathbf{x}} = \frac{\mathbf{h}}{||\mathbf{h}||}\mathbf{x}$, where $\mathbf{x}$ is the transmit signal from AAS, (11) reduces to:

$$y = \sqrt{P_t F 10^{\frac{G_m}{10}}} ||\mathbf{h}|| \mathbf{x} + \mathbf{n} \quad (13)$$

The downlink SNR, $\gamma$ for the UAV is then given as:

$$\gamma = \frac{P_t F 10^{\frac{G_m}{10}} ||\mathbf{h}||^2}{\sigma_n^2} \quad (14)$$

## III. MATHEMATICAL FRAMEWORK FOR RANGE EXTENSION

For range extension leveraging large number of antennas, we derive an analytical expression for the number of antennas, $M$ as a function of the range $d$, target SNR, $\gamma$ and height of UAV, $h$ by substituting (9) in (14):

$$\gamma = \frac{P_t F 10^{\frac{G_m}{10}}}{\sigma_n^2} \| [\mathbf{w}_1^T \sqrt{G} \mathbf{v}_1 \ldots \mathbf{w}_M^T \sqrt{G} \mathbf{v}_M] \|^2 \quad (15)$$

$$= \frac{P_t F 10^{\frac{G_m}{10}}}{\sigma_n^2} \text{Tr}\left(\begin{bmatrix} \mathbf{w}_1^T \sqrt{G} \mathbf{v}_1 \\ \vdots \\ \mathbf{w}_M^T \sqrt{G} \mathbf{v}_M \end{bmatrix} [\mathbf{w}_1^T \sqrt{G} \mathbf{v}_1 \ldots \mathbf{w}_M^T \sqrt{G} \mathbf{v}_M]\right)$$

$$= \frac{P_t F 10^{\frac{G_m}{10}}}{\sigma_n^2} \text{Tr}\begin{bmatrix} (\mathbf{w}_1^T \sqrt{G} \mathbf{v}_1)^2 & \ldots & \mathbf{w}_1 \mathbf{w}_M G \mathbf{v}_1 \mathbf{v}_M \\ \vdots & \ddots & \vdots \\ \mathbf{w}_M G \mathbf{v}_M \mathbf{w}_1^T \mathbf{v}_1 & \ldots & (\mathbf{w}_M^T \sqrt{G} \mathbf{v}_M)^2 \end{bmatrix}$$

$$= \frac{P_t F 10^{\frac{G_m}{10}}}{\sigma_n^2} \left((\mathbf{w}_1^T \sqrt{G} \mathbf{v}_1)^2 + \ldots + (\mathbf{w}_M^T \sqrt{G} \mathbf{v}_M)^2\right) \quad (16)$$

In this study, we assume $\phi_{scan} = 0$ as we are concerned with elevation beamforming only. Hence (16) reduces to:

$$\gamma = \frac{P_t F 10^{\frac{G_m}{10}}}{\sigma_n^2} M (\mathbf{w}_1^T \sqrt{G} \mathbf{v}_1)^2$$

$$= \frac{P_t F 10^{\frac{G_m}{10}} MG}{N\sigma_n^2} \left(\begin{bmatrix} \exp(-i2\pi(1-1)\frac{d_y}{\lambda}\cos\theta_{tilt}) \\ \vdots \\ \exp(-i2\pi(N-1)\frac{d_y}{\lambda}\cos\theta_{tilt}) \end{bmatrix}^T \begin{bmatrix} \exp(i2\pi(1-1)\frac{d_y}{\lambda}\cos\theta) \\ \vdots \\ \exp(i2\pi(N-1)\frac{d_y}{\lambda}\cos\theta) \end{bmatrix}\right)^2$$

$$= \frac{P_t F 10^{\frac{G_m}{10}} MG}{N\sigma_n^2} \underbrace{\left(\sum_{k=1}^{N} \exp(-i2\pi(k-1)\frac{d_y}{\lambda}\cos\theta_{tilt}) \exp(i2\pi(k-1)\frac{d_y}{\lambda}\cos\theta)\right)^2}_{A} \quad (17)$$

where Tr is the Trace operator. We further identify two cases of (17): tracking and no tracking. Tracking in airborne systems is the scenario when both $\theta$ and $\theta_{tilt}$ are aligned with each other as shown in Fig. 2, such that the UAV always receives main lobe gain as it moves. The second case of no tracking occurs when this type of dynamic beamforming is absent. Applying exponential sum formulas to (17), the term $A$ in (17) for both cases reduces to:

$$A = \begin{cases} N & \theta = \theta_{tilt} \\ \frac{\sin(0.5Ma)}{\sin(0.5a)} \exp\left(\frac{ia(N-1)}{2}\right) & \theta \neq \theta_{tilt} \end{cases} \quad (18)$$

where $a = 2\pi \frac{d_y}{\lambda}(\cos\theta - \cos\theta_{tilt})$.

From geometry in Fig. 1, $\theta$ can be expressed in terms of $h$ and $d$ as: $\theta = \cos^{-1}(h/d)$. Substituting this $\theta$ in (17)-(18) and then performing algebraic manipulation leads to the expression for number of antennas in terms of range, height and target SNR in dB ($\gamma_{db}$) given by (19), where $\lceil . \rceil$ is the ceiling function.

## IV. NUMERICAL RESULTS AND ANALYSIS

Fig. 3 shows the SNR in (17) as a function of number of antennas and range for $P_t = 30$ dBm, $f = 2.0$ GHz and $h = 1000$m. In contrast to prior works which depict SNR as a function of number of antennas only, our study adds another dimension of range into SNR expression. The effect of number of antennas in tandem with range can be observed for SNR in Fig. 3. Although SNR increases with increasing number of antennas, as expected range extension causes degradation in SNR.

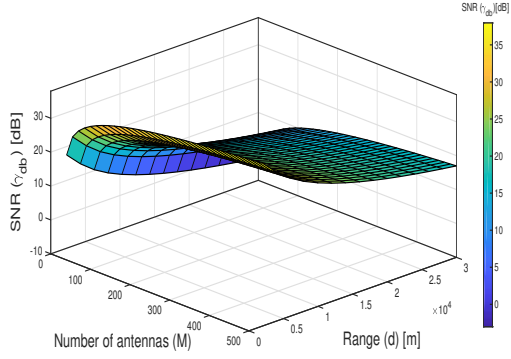

**Fig. 3:** SNR as a function of number of antennas and range (with tracking).

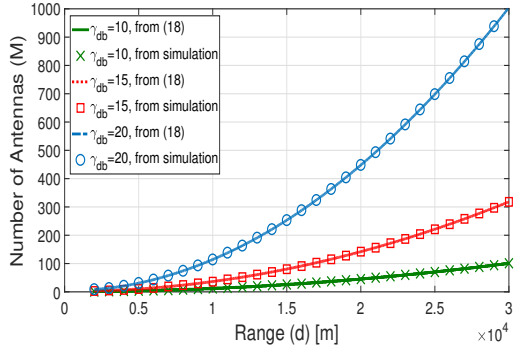

**Fig. 4:** Number of antennas verses range for different target SNRs at $h = 1000$m and $f = 2$GHz.

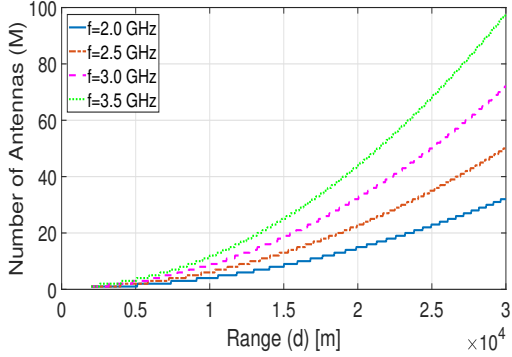

**Fig. 5:** Number of antennas verses range with different frequencies for target SNR of 5dB at $h = 1000$m.

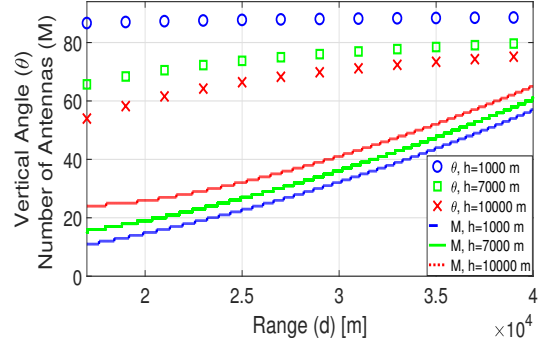

**Fig. 6:** Number of antennas and vertical angle verses range with different UAV heights for target SNR of 5dB at $f = 2$GHz.

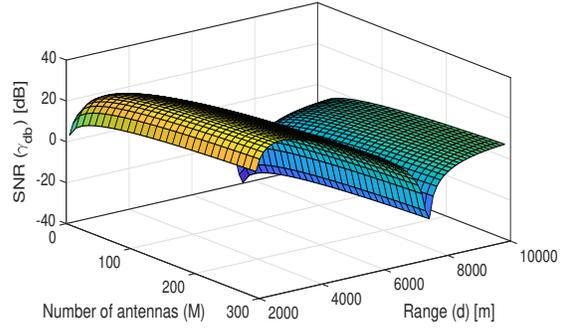

**Fig. 7:** SNR as a function of number of antennas and range at $\theta_{tilt} = 70^o$ (no tracking).

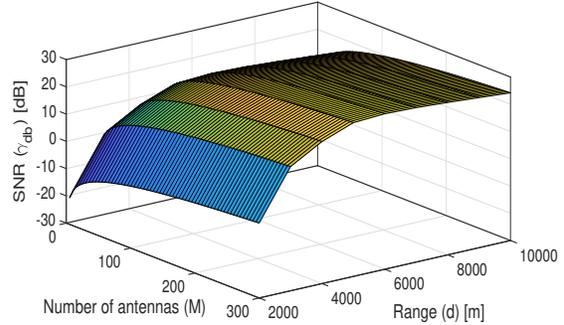

**Fig. 8:** SNR as a function of number of antennas and range at $\theta_{tilt} = 85^o$ (no tracking).

The proposed framework also allows to design the system for a target received signal strength or received SNR. More specifically, for given design constraints governed by the target SNR, our model allows to determine the number of antennas required to increase the range from $d$ to $md$, where $m$ can be any real number. This is depicted in Fig. 4. For higher target SNRs, number of antennas increases to achieve the same range extension. Our derived expression in (19) is also corroborated through simulations in Fig. 4.

Our proposed framework can be extended for any frequency and UAV heights. Fig. 5 quantifies the increase in number of antennas with increase in frequency, owing to the increased free space path loss at higher frequencies. The number of antennas also increase at higher UAV heights as shown in Fig. 6. Although the main lobe radiation pattern is in line with the tilt angle through beamforming or tracking, higher UAV heights result in higher vertical angles which in turn increase antenna attenuation in (2), leading to decrease in amplitude of main lobe from

$$M(\gamma_{db}, d, h) = \begin{cases} \left[\dfrac{10^{\frac{\gamma_{db}}{10}} 4\pi d \sigma_n^2}{NP_t \lambda^2 10^{\frac{G_m}{10}} 10^{-1.2\left(\left(\frac{\phi}{\phi_{3dB}}\right)^2 + \left(\frac{\cos^{-1}\left(\frac{h}{d}\right) - 90}{\theta_{3dB}}\right)^2\right)}}\right] & \theta = \theta_{tilt} \\[2em] \left[\dfrac{10^{\frac{\gamma_{db}}{10}} 4\pi d \sigma_n^2 N \sin(0.5a)}{\sin(0.5Na) P_t \lambda^2 10^{\frac{G_m}{10}} 10^{-1.2\left(\left(\frac{\phi}{\phi_{3dB}}\right)^2 + \left(\frac{\cos^{-1}\left(\frac{h}{d}\right) - 90}{\theta_{3dB}}\right)^2\right)}}\right] & \theta \neq \theta_{tilt} \end{cases} \quad (19)$$

(1). Hence, more antennas are required to compensate for the decreased antenna gain. Dynamically adjustable radiation patterns in an antenna array system can be realized through smart antenna techniques [13]-[15].

Next, we investigate the case when $\theta \neq \theta_{tilt}$ i.e., tracking-less scenario. Figures 7 and 8 show this case for two different values of $\theta_{tilt}$, $70^o$ and $85^o$. As compared to the case with tracking, $\theta$ is no longer in line with tilt angle of antenna as range varies. In Fig. 7, maximum SNR at around 3000m can be attributed the maximum value of antenna gain at 3000m, since a tilt angle of $70^o$ roughly corresponds to $\theta$ at a distance of 3000m. Similarly, in contrast to Fig. 3, SNR increases with range in Fig. 8 as the vertical angle increases with range and attains a maximum value where $\theta$ equals $\theta_{tilt} = 85^o$ (at $d = 10000$m).

## V. Conclusion

The problem of range extension in ground-to-air communications is addressed by leveraging large number of antennas at the base station. In contrast to previous works, where SNR as a function of number of antennas is computed at fixed distances or averaged over possible user positions, in this work, we add a new dimension of range to SNR derivation. The proposed mathematical framework allows dimensioning of systems with large number of antennas in terms of number of antennas for a given range and SNR threshold and vice versa. The framework also allows analysis of effect of different frequency bands and tracking and non-tracking scenarios. While results for tracking scenario provide quantification of intuitively expected trend where SNR increases with number of antennas and then saturates, non-tracking scenarios characterize counter-intuitive insights which can include both increasing and decreasing pattern depending on tilt angle. Thus the proposed framework provides a basic new tool for dimensioning systems with large number of antennas for ground-to-air communication, which can be extended to multiple users as well. Such investigations of multi-user interference using the proposed model can be focus of a future study.

## Acknowledgement

This work is supported by the National Science Foundation under Grant Number 1619346 and Grant Number 1718956.